# Multicriteria Evaluation and Sensitivity Analysis on Information Security

Irfan Syamsuddin
CAIR - Center for Applied ICT Research
Department of Computer and Networking Engineering
State Polytechnic of Ujung Pandang
Jl Perintis Kemerdekaan Km 10 Tamalanrea
Makassar 90245 INDONESIA

## ABSTRACT

Information security plays a significant role in recent information society. Increasing number and impact of cyber attacks on information assets have resulted the increasing awareness among managers that attack on information is actually attack on organization itself. Unfortunately, particular model for information security evaluation for management levels is still not well defined. In this study, decision analysis based on Ternary Analytic Hierarchy Process (T-AHP) is proposed as a novel model to aid managers who responsible in making strategic evaluation related to information security issues. In addition, sensitivity analysis is applied to extend our analysis by using several "what-if" scenarios in order to measure the consistency of the final evaluation. Finally, we conclude that the final evaluation made by managers has a significant consistency shown by sensitivity analysis results.

## 1. INTRODUCTION

The role of information security has gained more serious concerns particularly in the last two decades. The more dependent organizations on access, store, and transfer their information through the internet, the more probability of cyber security attacks they could face. In this regards, many cyber crime cases with significant financial losses have been periodically announced by several organizations. As a result, security managers are highly requested to be aware of the increasing security challenges which may occur anytime.

In fact, only few academic papers examine how security managers formulate strategic decision in such dynamic situations. This is because most organizations tend to hide the actual incidences since it might affect the image of organization. In addition, such announcements may damage public trust to the company because it is likely to reveal internal weaknesses that should not appear publicly (Fulford and Doherty, 2003). Therefore, empirical studies in this field are still open both from academic and professional to give constructive contributions. This study aims to fill the gap in information security literature particularly from managerial perspective. It is believed that by having an adequate model that could aid managers in evaluating information security issues, many potential damages related to cyber security attacks can be avoided or at least reduced in terms of its impact on organizations. Therefore, this paper is aimed at applying an evaluation framework of information security based on our previous work in 2009. In addition, the evaluation results will be also extended by sensitivity analysis to measure how consistent the final evaluation made by managers.

## 2. LITERATURE REVIEW

In evaluating information security there are several perspectives that should be involved in order to produce a better and more effective information security implementations in the future (Wylder, 2004). The elements encompass people, systems, information and procedure with respect to security and privacy issues.

In this regard, many perspectives should be considered in evaluation information security as argued by Syamsuddin (2012) considering complexity of recent security breaches in organization that do not only involves technical issues but also has non technical ones such as economic, managerial as well as cultural effects. Householder, et.al (2002) mention historical technical security issues from hardware and applications security until computer network security, wireless security and internet or cyber security. These all justify the root of current cyber security issues are basically from technical perspectives (Syamsuddin, 2012). However, along with advancement of information technology and reliance of business and government organizations on information, cyber security is no longer a technical domain.Awareness on economic impact of information security has just discussed in the last decade. Anderson (2001) is among the earliest researcher who describe relationship between information security and economy by proposing a theory called the economics of information security. Under this concept, various economic mechanisms are applied to analyze cyber security behavior such as security incentives, investment and financial information sharing (Gordon and Loeb, 2002).

Similarly, managerial aspect of information security plays more and more important role just several years ago as it plays an essential role in ensuring information handling within organization (Filipek, 2007). Weakness in managerial handling of information security may result in serious damage to information resources of an organization.

Successful information security could not be achieved unless it has become a daily life of people within an organization. In this regards, cultural information security is believed as a fundamental solution to any kind of technical security applied in an organization. Without strong cultural approach any security technologies will not work properly. Lack of cultural awareness of security in an organization is cited as the source of a number security breaches (Martins and Eloff, 2002). Education and reward-punishment method are promoted to cultivate security culture at organizational level (Thomson and von Solms, 1998).

Syamsuddin and Hwang (2009) justify that in evaluating information security, one should look at the problem from four perspectives above namely, technology, management, economy and culture. They point out that further decision or evaluation should consider three main aspects of security objectives called CIA which stands for confidentiality, integrity and availability.

Bacik (2008) describes confidentiality integrity and availability as follows. Confidentiality reflects protection of the privacy users in respect to their own information. It is the property of preventing disclosure of information to unauthorized individuals or systems.

Integrity is the property of preventing any possible changes of information. It means by ensuring the integrity, information or data cannot be modified or edited without authorization of the owner. In other words, integrity keeps the intact of data and that only authorized user able to access or modify it.

Availability is the property of providing appropriate information when required. It means that for any information system to serve its purpose, the information must be available when it is needed. Availability ensures the computing systems used to store and process the information, the security controls used to protect it, and the communication channels used to access it must be functioning correctly.

## 3. MULTICRITERIA EVALUATION

### 3.1 Analytic Hierarchy Process

One of the natures of evaluation is the existence of multiple alternatives to be chosen and multiple aspects or criteria to assist evaluation processes. Multicriteria Evaluation (MCE) or often called Multicriteria Decision Analysis (MCDA) or Multicriteria Decision Making (MCDM) is a method that satisfies the need to incorporate multiple criteria and alternatives at the same time under equal judgment (Ertay, et.al, 2012).

Among several methodology of MCE, Analytic Hierarchy Process developed by Saaty (1980) is one that widely accepted and applied by researchers and practitioners from various disciplines.

To date, thousands of AHP applications can be seen in business, management, government, military and many other areas where multi criteria decision problems exist. The strength of AHP also lies on its simplistic mathematical calculation to perform decision making processes. In addition, both qualitative and quantitative analysis can be done simultaneously with AHP which is rarely found in other decision making methods (Vadya and Kumar, 2006). Details of logical algorithm behind AHP method might be read directly from Saaty (1980) while example of mathematical application with OpenCalc (an open source software) might be read from Syamsuddin and Hwang (2010).

Like other fields, AHP is also not immune from criticism that reveals some of its weaknesses. There is no single method On the other hand, several limitations addressed to AHP by many researchers. Preserving consistency is the most challenging effort in conducting an AHP survey. If the consistency ratio (CR) is more than 0.1, than respondents are required to review their judgments until the minimum standard of CR is satisfied.

### 3.2 Ternary AHP

In reality, human being can easily compare one to other simply by saying "*A better than B*" or "*A worse than B*". This situation was adopted by Takahashi (1990) who apply modified AHP in sports game. According to Takahashi (1990), in reality there are three possible conditions in sports game namely win, lose or draw. Readers may refer to Takahashi's paper (1990) for details argument and mathematical foundation behind Ternary AHP.

In short, instead of using classical AHP of Saaty (1980), we prefer to apply Takahashi's Ternary Analytic Hierarchy Process (1990) since it adequately meets main requirements for this study. Moreover, the decision makers require lesser time and put minimal efforts while significantly reducing possibility of inconsistency ratio. Unlike classical AHP which employs 1 to 9 scales, T-AHP uses only three values to represent one's preference or judgment. As can be seen in Table 1 below

**Table 1. Ternary AHP's preference values**

| No | Preference Values | Description |
|---|---|---|
| 1 | 1 | equally important between criteria/alternative $i$ to $j$ |
| 2 | $\theta$ | criteria/alternative $i$ is more important than $j$ |
| 3 | $1/\theta$ | reciprocal state of number 2 |

One of the main advantages of applying ternary numbers as described above is significantly reduce potential judgment conflicts which eventually lead to better consistency ratio. This advantage improved by Takeda (2001) who justify potential applications of Ternary AHP in uncertainty and indetermination and incomplete certain information.

Besides, Nishizawa and Takahashi's paper (2007) that illustrate applicability of Ternary AHP in stochastic models, minimax as well as least square estimation methods adds a series of benefit of Ternary AHP.

## 4. ANALYSIS and DISCUSSION

### 4.1 Evaluation Model

The proposed model for strategic information security decision analysis is represented in figure 4. It is structured into four levels of hierarchy consisting of goal, criteria, sub-criteria and alternatives. The first level of goal represents "Information Security Evaluation" as the aim of this study. The second level of criteria consists of four items, namely

Management (M), Technology (T), Economy (E) and Culture (C) which represent the four main aspects of information security. Subsequently, the sub criteria level consists of ten additional criteria grouped by each main criterion previously.

In the last layer, three alternatives are given. They are Integrity (In) and Availability (Av) which represent strategic solution for future information security implementation. Typical survey question at second layer is exemplified like *"Which one is more important in Information Security Evaluation, Technical criteria or Management criteria?"* There are only three possible answer to choose, 1 for equally important, θ for Technical is more important than Managerial and 1/θ for in contrary of second option of Managerial is more important than Technical.

Finally after performing pairwise comparison at all levels, results for main criteria and alternative are gained. Cultural aspect in terms of security education and reward/ punishment approaches is found to be the main focus for future strategic information security which accounted for 0.409. It is followed by managerial improvement of 0.241. Technical and economical perspectives seem to have similar weight of 0.175.

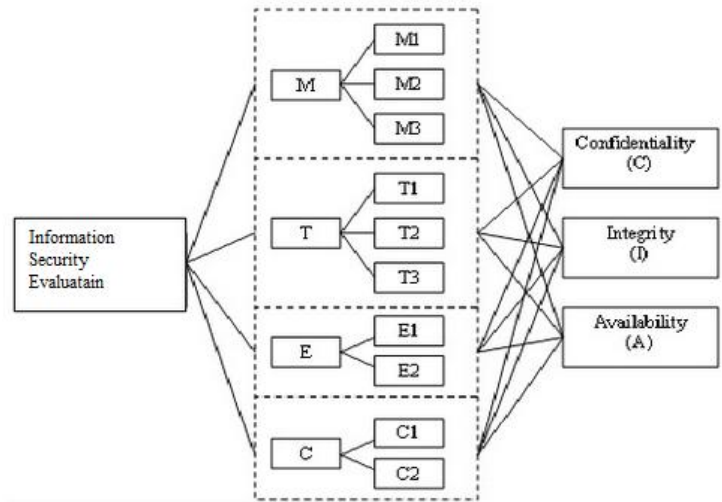

**Fig. 1. Decision Analysis Model**

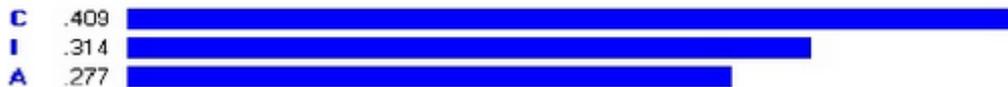

**Fig. 2. Final evaluation**

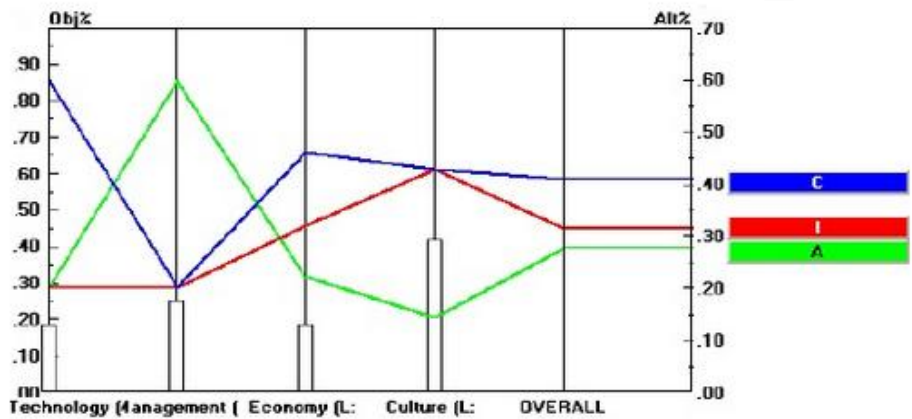

**Fig. 3. Simulation final evaluation with sensitivity analysis**

Figure 2 reveals the final preference of alternative with respect to the goal that shows the best evaluation of decision maker of CIA (confidentiality, integrity and availability) resource allocation in this study. Confidentiality is accounted on the top preference with 0.409 followed by integrity and availability both with 0.314 and 0.277 respectively.

The findings indicate that most efforts should be given more on improving the confidentiality of data and information systems as a key strategy for the future. On the other hand, efforts to ensure integrity should also get adequate attentions for the future strategic information security programs, while availability of data and information systems are recommended with lesser concerns due to its maturity in its development.

## 4.2 Sensitivity Analysis

Applying sensitivity analysis to such decision making processes is essential to ensure the consistency of final decision. Through sensitivity analysis, different "what-if" scenarios can be visualized which are helpful to observe the impact of changing on criteria to final alternative rank. Sensitivity analysis as shown in figure 3 lets evaluator to observe how final evaluation is likely to change. It also helps in measuring how much changes made by certain extent of deviations in weights of criteria.

In this case, simulation of sensitivity analysis is carried out by making gradual changes on values of each criterion, whether technology (T), management (M), economy (E) or culture (C), and then observing the rank order due to such changes. It is revealed that by shifting the value of each criterion lowering down to zero point, it did not have any effect would not result in any changes to the first rank (confidentiality).

Rank reversal occurs only to the second and third ranks (integrity and availability respectively) when cultural aspect is reduced to zero point. Only in this particular, rank reversal occurs when availability jump on top over integrity. Overall, based on sensitivity analysis, it can be concluded that the final decision is consistent and reliable.

## 5. CONCLUSION

The application of Ternary AHP is sound and fit to the case of evaluation of information security. Its simplicity has assisted evaluator to reach high level consistency level and also reduce possibility of rank reversal as demonstrated in sensitivity analysis through various simulations.

In short, the evaluation suggest the essential role of cultural approaches such as security education or training and reward punishment practices in improving information security program in the organization. Additionally, a strategic information security in the future must put more concerns on confidentiality issues of data and information systems, before integrity and availability.

## 6. ACKNOWLEDGMENTS

I would like to five special thanks to Politeknik Negeri Ujung Pandang and Center for Applied ICT Research for supporting this study and finalizing the publication.